\documentclass[20pt]{article}
\usepackage{amsmath}
\usepackage{amssymb}
\usepackage{latexsym}
\usepackage{amsmath}
\usepackage{lineno}
\usepackage{graphicx}
\usepackage{float}
\usepackage{subfigure}
\usepackage{geometry}
\usepackage{graphics}
\usepackage{subfigure}
\usepackage{graphicx}
\usepackage{multirow}
\usepackage{booktabs}
\usepackage{hyperref}
\usepackage{cite} 
\makeatletter

\newcommand{\be}{\begin{equation}}
\newcommand{\ee}{\end{equation}}

\begin{document}
\begin{center}
\large {\bf The Effects of Minimal Length, Maximal Momentum and Minimal Momentum in Entropic Force}
\end{center}

\begin{center}
Zhong-Wen Feng $^1$
 $\footnote{E-mail:  zwfengphy@163.com}$
Shu-Zheng Yang $^2$
 $\footnote{E-mail:  szyangcwnu@126.com}$
Hui-Ling Li $ ^{1, 3}$
 $\footnote{E-mail:  LHL51759@126.com}$
 Xiao-Tao Zu ${^1}$
 $\footnote{E-mail:  xtzu@uestc.edu.cn}$
\end{center}

\begin{center}
\textit{1. School of Physical Electronics, University of Electronic Science and Technology of China, Chengdu, 610054, China\\
2. Department of Astronomy, China West Normal University, Nanchong, 637009, China\\
3. College of Physics Science and Technology, Shenyang Normal University, Shenyang, 110034, China}
\end{center}

\noindent
{\bf Abstract:} In this paper, the modified entropic force law is studied by using a new kind of generalized uncertainty principle which contains a minimal length, a minimal momentum and a maximal momentum.  Firstly, the quantum corrections to the thermodynamics of a black hole is investigated. Then, according to Verlinde's theory, the generalized uncertainty principle (GUP) corrected entropic force is obtained. The result shows that the GUP corrected entropic force is related not only to the properties of the black holes, but also to the Planck length and the dimensionless constants \(\alpha_0\) and \(\beta_0\). Moreover, based on the GUP corrected entropic force, we also derive the modified Einstein's field equation (EFE) and the modified Friedmann equation.

\section{Introduction}
The existence of thermodynamics of black holes is a great discovery for the foundations of physics. \cite{ch1,ch2,ch3,ch4,ch5,ch5xxx,ch6,ch7,ch8,ch8xxx,ch9xxx}. This idea was proposed by Bekenstein who proved the entropy of a black hole is  $S = {{Ak_B c^3 } \mathord{\left/ {\vphantom {{Ak_B c^3 } {4\hbar G}}} \right. \kern-\nulldelimiterspace} {4\hbar G}}$, where  $A$ is the horizon area,  $k_B$ is the Boltzmann constant, $\hbar$ is the Planck constant and $G$ is the Newton's gravitational constant, respectively \cite{ch2}. Later, based on the entropy of black hole, Hawking showed that the Schwarzschild (SC) black hole emits thermal radiation, the temperature of SC black hole is proportional to the surface gravity on the event horizon $\kappa$ , namely,  $T = {\kappa  \mathord{\left/ {\vphantom {\kappa  {2\pi }}} \right. \kern-\nulldelimiterspace} {2\pi }}$ \cite{ch5,ch5xxx}. Furthermore, in \cite{ch8,ch8xxx}, the authors proved that
the entropy and the temperature of black hole satisfy the laws of thermodynamics. These discoveries indicate that the thermodynamics of black hole have profound connection with the gravity.

In order to investigate the deeper-seated relation between thermodynamics and the gravity, Jacobson assumed the spacetime as a kind of gas, its entropy is proportional to the area, then using the fundamental Clausius relation and the equivalence principle, he demonstrated that the Einstein field equation is nothing but an state equation of this kind of gas. Following this agreement, the Einstein field equation can be derived from the first law of thermodynamics together with the relation between the entropy and the horizon area of a black hole \cite{ch9}. Subsequently, Padmanabhan pointed out that, in the static spherically symmetric spacetimes, the gravitational field equations on the horizon can be rewritten as a form of the ordinary first law of thermodynamics \cite{ch10}. Inspired by Padmanabhan's idea, people found that the Einstein's equation is a thermodynamic identity. Those works explain why the field equations should encode information of horizon thermodynamics \cite{ch11,ch12,ch13,ch14,ch15,ch16}.

In 2011, Verlinde proposed a remarkable new perspective on the relation between the gravity and the thermodynamics. Based on Sakharov's idea \cite{ch17} and the holographic principle, Verlinde pointed out that the gravity is no longer a fundamental force, instead, it can be explained as an entropic force which arises from the change of information when material bodies move away from the screens of holographic systems \cite{ch18}. In his paper, Verlinde showed various interesting results. For example, the the second law of Newton can be obtained by incorporating the entropic force with the Unruh temperature. Using the entropic force together with the holographic principle and the equipartition law of energy, one can yield the Newton's law of gravitation. Moreover, an astounding discovery should be mentioned that the Einstein's equation can be derived from the theory of entropic force. This new proposal of the gravity has received wide attention causing people to do many relevant works to discuss the entropic force \cite{ch23,ch21,ch22,ch1c+}.

On the other hand, a lot of works showed that the original thermodynamics of black holes would not be held when considering the quantum gravity effects \cite{ch1m+,ch2m+,ch3m+,ch673+,chb4+,ch4m+}. Various theories of quantum gravity suggest that the existence of a minimal observable length, which can be identified with the Planck scale. This view is supported by many Gedanken experiments and is applied to different physical systems \cite{ch1d+,ch2d+,ch3d+,ch4d+}. The generalized uncertainty principle (GUP) is one of the most important theories, which is modified by the minimum measurable length. The applications of GUP has been widely studied \cite{ch1e+,ch2e+,ch3e+,ch1a+,ch2a+,ch3a+,ch4a+,ch5a+,ch25,ch618a+}. Especially, the GUP effect on the micro black holes have been deeply discussed in \cite{ch1t+,ch2t+,ch3t+}. In \cite{ch19}, combining the GUP with the thermodynamics of black holes, the authors investigated the modified Hawking temperature and the entropy; their results showed that the GUP corrected thermodynamics of black holes is different from that of the original cases. The GUP can stop the Hawking radiation in the final stages of black holes' evolution and lead to the remnants of black holes. Therefore, the GUP is considered as a good tool to solve the information paradox problem of black holes. In \cite{ch24,ch26,ch26+,ch27,ch28,chb1+,chb2+,chb3+}, people have investigated the GUP corrected entropic force by using the GUP corrected thermodynamics. The implications of modified entropic force have been investigated in many contexts such as cosmology \cite{ch24,ch26,ch26+,ch27}, Hamiltonians of the quantum systems in quasi-space \cite{ch28}, quantum walk \cite{chb1+} and quarkonium binding \cite{chb2+,chb3+}. In fact, the expression of GUP is not unique. However, in most of the papers, the expression of GUP is limited to two forms. One form only has a minimal length  $\Delta x\Delta p \ge \hbar \left[ {1 + \alpha _0^2 \ell _p^2 \left( {\Delta p} \right)^2 } \right]$ \cite{chb4+}. The other form has a minimal length and a maximal momentum $\Delta x\Delta p \ge \hbar \left[ {1 - \alpha _0 \ell _p \Delta p + \alpha _0^2 \ell _p^2 \left( {\Delta p} \right)^2 } \right]$ \cite{chb5+}. According to the previous works, it is interesting to raise the question whether it is possible to derive a general form of GUP which contains a minimal length, a minimal momentum and a maximal momentum? Actually, in \cite{chb6+}, the authors pointed out that the existence of a minimal length $\Delta x_{\min }$ comes from the fact that a string cannot probe its distances smaller than the length, and the maximal momentum $\Delta p_{\max }$ originated in the doubly special relativity (DSR) that predicts there exists an upper bound for the momentum of a particle. For the minimal momentum $\Delta p_{\min }$, it is well known that the notion of a plane wave does not appear in the a general curved spacetime; hence, it indicates that there exists a limit to the precision with which the corresponding momentum can be described. According to this phenomenon, one can express a nonzero minimal uncertainty in momentum measurement, that is $\Delta p_{\min }$. Taking these facts into account, Nozari and Saghaf introduced the most general form of GUP, which admits a minimal length, a minimal momentum and a maximal momentum. The GUP is given by
\begin{equation}
\label{eq1}
\Delta x\Delta p \ge \hbar \left[ {1 - \alpha _0 \ell _p \Delta p + \alpha _0^2 \ell _p^2 \left( {\Delta p} \right)^2  + \beta _0^2 \ell _p^2 \left( {\Delta x} \right)^2 } \right],
\end{equation}
where $\Delta x$ and $\Delta p$ are the uncertainties for position and momentum, the  $\alpha _0$  and  $\beta _0$  are dimensionless constants of the order of unity that depends on the details of the quantum gravity hypothesis, $\ell _p  = \sqrt {{{\hbar G} \mathord{\left/ {\vphantom {{\hbar G} {c^3 }}} \right. \kern-\nulldelimiterspace} {c^3 }}}  \approx 10^{ - 35} m$  is the Planck length, respectively \cite{ch29}. From Eq. (\ref{eq1}), one can easily obtain the minimal length  $\Delta x_{\min }  = \alpha_0 \ell _p$, the minimal momentum  $\Delta p_{\min }  = 2\beta _0 \ell _p$ and the maximal momentum  $\Delta p_{\max }  = {{\left[ {1 + \sqrt {1 - \left( {1 + \alpha _0^2 \beta _0^2 \ell _p^4 } \right)} } \right]} \mathord{\left/ {\vphantom {{\left[ {1 + \sqrt {1 - \left( {1 + \alpha _0^2 \beta _0^2 \ell _p^4 } \right)} } \right]} {\alpha_0 \ell _p }}} \right.  \kern-\nulldelimiterspace} {\alpha_0 \ell _p }}$. Under the standard limit, that is  $\Delta x \gg \ell _p$, the Eq. (\ref{eq1}) becomes the Heisenberg uncertainty principle (HUP). It is well known that GUP has a great effect on the thermodynamics of black holes and the entropic force, that would bring many new results. Therefore, in this paper, using Eq. (\ref{eq1}), we first investigate the modified thermodynamics of a black hole. Then, following Verlinde's viewpoint, the modified number of bits  $N$ is obtained. The modified number of bits $N$ leads to the GUP corrected entropic force. Considering the GUP corrected entropic force, the gravitational force, the gravitational potential, the Einstein's field equation and the Friedmann equation are modified.

This paper is organized as follows. The quantum  corrections to the entropy and Hawking temperature is derived in section 2. According to Verlinde's theory, the GUP corrected entropic force is calculated in section 3. In section 4 and section 5, using the GUP corrected entropic force, the modified Einstein's field equation and the modified Friedmann equation is investigated.  The last section is devoted to the conclusions.

\section{The GUP Impact on The Thermodynamics of A Black Hole}
In order to calculate the GUP impact on the thermodynamics of a black hole, one needs to solve Eq. (\ref{eq1}) as a quadratic equation in  $\Delta p$. The result is given by
\begin{equation}
\label{eq2}
\Delta p \ge \frac{{\Delta x + \hbar \alpha _0 \ell _p }}{{2\hbar \alpha _0^2 \ell _p^2 }}\left\{ {1 - \sqrt {1 - \frac{{4\hbar ^2 \alpha _0^2 \ell _p^2 \left[ {1 + \beta _0^2 \left( {\Delta x} \right)^2 \ell _p^2 } \right]}}{{\left( {\Delta x + \hbar \alpha _0 \ell _p } \right)^2 }}} } \right\},
\end{equation}
where we choose the negative-signed solution since only it can recover the HUP in the classical limit \( {\ell _p }\rightarrow 0  \). Employing the Taylor expansion, Eq. (\ref{eq2}) can be rewritten as
\begin{equation}
\label{eq3}
\Delta p \ge \frac{1}{{\Delta x + \alpha _0 \ell _p }}\left\{ {1 + \left[ {\beta _0^2 \left( {\Delta x} \right)^2  + \frac{{\alpha_0 ^2 }}{{\left( {\Delta x + \alpha _0 \ell _p } \right)^2 }}} \right]\ell _p^2 + {\cal O}\left( {\ell _p^4 } \right)} \right\},
\end{equation}
where we are setting $\hbar=1$. As \cite{ch19,ch30} have pointed out, the uncertainty momentum $\Delta p$  can be defined as the energy $\omega$  of emitted photon from black hole. Therefore, based on Eq. (\ref{eq3}), the low bound for the energy is
\begin{equation}
\label{eq4}
\omega \ge  \frac{1}{{\Delta x + \alpha _0 \ell _p }}\left\{ {1 + \left[ {\beta _0^2 \left( {\Delta x} \right)^2  + \frac{{\alpha _0^2 }}{{\left( {\Delta x + \alpha _0 \ell _p } \right)^2 }}} \right]\ell _p^2} +{\cal O}\left( {\ell _p^4 } \right) \right\}.
\end{equation}
Next, assuming an emitted particle with energy $\omega$  and size $R$, for any black hole absorbing or emitting this particle, the minimal change in the horizon area of a black hole can be expressed as
\begin{equation}
\label{eq5}
\Delta A_{\min }  \ge 8\pi \omega R\ell _p^2 .
\end{equation}
In the classical model, people can set  $R=0$. However, according to the arguments of \cite{ch2}, the size of a quantum particle cannot be smaller than  $\Delta x$. Therefore, it implies the existence of a finite boundary given by
\begin{equation}
\label{eq6}
\Delta A_{\min }  \ge 8\pi \omega \Delta x\ell _p^2 .
\end{equation}
By substituting Eq. (\ref{eq4}) into the inequality (\ref{eq6}), then, using the relation of minimal length  $\Delta x_{\min }  = \alpha _0 \ell _p$, the boundary turns out to be
\begin{equation}
\label{eq7}
\Delta A_{\min }  \ge 4\pi \ell _p^2 \left\{ {1 + \left[ {\beta _0^2 \left( {\Delta x} \right)^2  + \frac{{\alpha _0^2 }}{{\left( {2\Delta x} \right)^2 }}} \right]\ell _p^2 } + {\cal O}\left( {\ell _p^4 } \right) \right\}.
\end{equation}
Now, considering the case that a photon is emitted by the SC black hole \cite{ch19,ch30,ch31}. Near the event horizon of SC black hole, the position uncertainty of a photon is the order of the radius of the black hole, that is, $\Delta x = 2r_s$, where $r_s$ is the radius of SC black hole. According to the area of the SC black hole  $A = 4\pi r_s^2$, the relation between $\Delta x$  and $A$  can be expressed as  $\left( {\Delta x} \right)^2  = 4r_s^2  = {A \mathord{\left/ {\vphantom {A \pi }} \right. \kern-\nulldelimiterspace} \pi }$. Substituting this relation into Eq. (\ref{eq7}), the above equation can be rewritten as the following expression
\begin{equation}
\label{eq8}
\Delta A_{\min }  \simeq \lambda \ell _p^2 {\rm{ }}\left[ {1 + \left( {\beta _0^2 \frac{A}{\pi } + \frac{{\pi \alpha _0^2 }}{{4A}}} \right)\ell _p^2 } + {\cal O}\left( {\ell _p^4 } \right) \right]{\rm{ }},
\end{equation}
with  $\lambda$ being a undetermined coefficient that is greater than  $4 \pi$. In the previous works, people proved that the entropy of black holes depended on the area of horizon. Moreover, the ideas of information theory also showed the minimal increase of entropy is conjectured related to the value of the area  $A$. According to \cite{ch2,ch32}, the fundamental unit of entropy as one bit of information can be denoted as  $\Delta S_{\min }  = b = \ln 2$, so that one can easy obtain
\begin{equation}
\label{eq9}
\frac{{dS}}{{dA}} = \frac{{\Delta S_{\min } }}{{\Delta A_{\min } }} = \frac{b}{{\lambda \ell _p^2 }}\left[ {1 + \left( {\beta _0^2 \frac{A}{\pi } + \frac{{\pi \alpha _0^2 }}{{4A}}} \right)\ell _p^2 } + {\cal O}\left( {\ell _p^4 } \right) \right]^{ - 1} .
\end{equation}
In accordance with the idea of entropy-area law, we obtain a constant ${b \mathord{\left/ {\vphantom {b \lambda }} \right. \kern-\nulldelimiterspace} \lambda } = {{k_B } \mathord{\left/ {\vphantom {{k_B } 4}} \right. \kern-\nulldelimiterspace} 4}$. Putting this constant into the above equation and expanding it, then integrating the result, the GUP corrected entropy is obtained as follows
\begin{equation}
\label{eq10}
S = \frac{{Ak_B }}{{4\ell _p^2 }}\left\{ {1 - \left[ {\frac{{\alpha _0^2 \pi }}{{4A}}{\rm{ln}}\left( {\frac{A}{{4\ell _p^2 }}} \right) + \frac{{A\beta _0^2 }}{{2\pi }}} \right]\ell _p^2  + {\cal O}\left( {\ell _p^4 } \right)} \right\}.
\end{equation}
It is clear that the GUP corrected entropy is proportional to the area of horizon  $A$, Planck length  $\ell _p$, and the dimensionless constants  $\alpha_0$ and  $\beta_0$. When ignoring the $\alpha_0$  and  $\beta_0$, Eq. (\ref{eq10}) reduces to the original entropy of the black hole. Moreover, it can be found that the first correction term of Eq. (\ref{eq10}) is logarithmic in the  $A$,  $\ell _p$,  $\alpha_0$, which is coincident with previous findings \cite{ch33,ch34,ch35}. It should be noted that the second correction term is proportional to $\beta_0$ goes like $A^2$, if $\beta _0 \geq \sqrt {{{2\pi } \mathord{\left/ {\vphantom {{2\pi } A}} \right. \kern-\nulldelimiterspace} A}}$, the second correction term is in principle larger than the leading term proportional to $A$. In order to avoid this paradoxical situation, it requires that $\beta _0  < \sqrt {{{2\pi } \mathord{\left/ {\vphantom {{2\pi } A}} \right. \kern-\nulldelimiterspace} A}}$. Meanwhile, one can calculate the GUP corrected Hawking temperature based on Eq. (\ref{eq10})
\begin{equation}
\label{eq11}
T = \frac{\kappa }{{8\pi }}\frac{{dA}}{{dS}} = \frac{\kappa }{{2\pi k_B }}\ell _p^2 \left[ {1 + \left( {\frac{{A\beta _0^2 }}{\pi } + \frac{{\pi \alpha _0^2 }}{{4A}}} \right)\ell _p^2  + {\cal O}\left( {\ell _p^4 } \right)} \right],
\end{equation}
where $\kappa$  is the surface gravity of black holes. For the SC black hole, one sets  $\kappa  = {1 \mathord{\left/ {\vphantom {1 {4M}}} \right. \kern-\nulldelimiterspace} {4M}}$, if setting  $\alpha _0  = \beta _0  = 0$, the GUP corrected temperature reduces to the original Hawking temperature.

\section{The Modified Newton's Law of Gravitation Due to The GUP}
In this section, we will investigate the GUP impact on the Newton's law of gravitation. For revealing the entropic force, Verlinde used the holographic principle and the first law of thermodynamics. When a test particle approaches a holographic screen, the entropic force of a gravitational system is expressed as
\begin{equation}
\label{eq12}
F\Delta x = T\Delta S,
\end{equation}
where $F$ is the entropic force, $T$  and $\Delta S$  are the temperature and the change of entropy on holographic screen,  $\Delta x$ is the displacement of the particle from the holographic screen,  respectively \cite{ch18}. The Eq. (\ref{eq12}) implies a non-zero force is proportional to a non-zero acceleration. Using the argument of Bekenstein, that is the change of entropy associated with the information on the boundary $\Delta S = 2\pi k_B$, it is found that the change in the entropy near the holographic screen is linear in the  $\Delta x$
\begin{equation}
\label{eq13}
\Delta S = {{2\pi k_B mc\Delta x} \mathord{\left/
 {\vphantom {{2\pi k_B mc\Delta x} \hbar }} \right.
 \kern-\nulldelimiterspace} \hbar },
\end{equation}
where  $\Delta x = {{\hbar} \mathord{\left/ {\vphantom {{\hbar } m c} } \right. \kern-\nulldelimiterspace} m c}$,  $m$ and $\hbar$ are the mass of elementary component and the Planck constant, respectively. Eq. (\ref{eq13}) is reminiscent of the osmosis across a semi-permeable membrane. Meanwhile, it should be noted that the $\Delta S$ is proportional to the mass of the elementary component. In order to understand this idea, one can postulate that a particle near the holographic screen is made up of two or more sub-particles and each sub-particle leads to the associated change in entropy after a displacement. Because the mass of the elementary component and the entropy are additive, it leads to that the $\Delta S$ is proportional to the $m$. Based on the concludes in \cite{ch36}, the horizon of black holes can be taken as a storage device for information. If one denotes the number of information by $N$ bits, the information is proportional to the area  $N = {{Ac^3 } \mathord{\left/ {\vphantom {{Ac^3 } G}} \right. \kern-\nulldelimiterspace} G}\hbar$. With the help of the entropy-area law $S = {{Ak_B c^3 } \mathord{\left/ {\vphantom {{Ak_B c^3 } {4\hbar G}}} \right. \kern-\nulldelimiterspace} {4\hbar G}}$, the number of bits obeys the following relation
\begin{equation}
\label{eq14}
N = {{4S} \mathord{\left/
 {\vphantom {{4S} {k_B }}} \right.
 \kern-\nulldelimiterspace} {k_B }}.
\end{equation}
Obviously, the number of bits is proportional to the entropy. Substituting Eq. (\ref{eq10}) into Eq. (\ref{eq14}), the number of bits is modified as follows
\begin{eqnarray}
\label{eq15}
N & = &\frac{A}{{\ell _p^2 }}\left\{ {1 - \left[ {\frac{{\alpha _0^2 \pi }}{{4A}}{\rm{ln}}\left( {\frac{A}{{4\ell _p^2 }}} \right) + \frac{{A\beta _0^2 }}{{2\pi }}} \right]\ell _p^2  + \mathcal{O}\left( {\ell _p^4 } \right)} \right\}
\nonumber \\
 &=& \frac{{Ac^3 }}{{G\hbar }}\left\{ {1 - \left[ {\frac{{\alpha _0^2 \pi }}{{4A}}{\rm{ln}}\left( {\frac{A}{{4\ell _p^2 }}} \right) + \frac{{A\beta _0^2 }}{{2\pi }}} \right]\ell _p^2  + \mathcal{O}\left( {\ell _p^4 } \right)} \right\},
\end{eqnarray}
where $\ell _p^2  = {{G \hbar} \mathord{\left/ {\vphantom {{Gh} {c^3 }}} \right. \kern-\nulldelimiterspace} {c^3 }}$. By setting the total energy of the black hole (or holographic system) is  $E$, and noting that the energy is divided evenly over the bits  $N$, it is easy to obtain that each bit carries an energy equal to  ${{k_B T} \mathord{\left/ {\vphantom {{k_B T} 2}} \right. \kern-\nulldelimiterspace} 2}$. According to the equipartition rule, the total energy can be expressed as
\begin{equation}
\label{eq16}
E = {{k_B NT} \mathord{\left/
 {\vphantom {{k_B NT} 2}} \right.
 \kern-\nulldelimiterspace} 2}.
\end{equation}
Using the relation $E = Mc^2$, then putting Eq. (\ref{eq12}) and Eq. (\ref{eq13}) into the above equation, one gets
\begin{equation}
\label{eq17}
F = {{4\pi c^3 Mm} \mathord{\left/
 {\vphantom {{4\pi c^2 Mm} {\hbar N}}} \right.
 \kern-\nulldelimiterspace} {\hbar N}}.
\end{equation}
Next, substituting  $N$ from Eq. (\ref{eq15}) into Eq. (\ref{eq17}), the GUP corrected Newton's law of gravitation becomes
\begin{equation}
\label{eq18}
F =  \frac{{GMm}}{{R^2 }}\left\{ {1 + \left[ {\frac{{\alpha _0^2 }}{{16R}}{\rm{ln}}\left( {\frac{{\pi R}}{{\ell _p^2 }}} \right) + 2\beta _0^2 R} \right]\ell _p^2 } + {\cal O}\left( {\ell _p^4 } \right) \right\}.
\end{equation}
In the above equation, one has $A=4 \pi R^2$. It is well known that the Newtonian gravitational force dominates at large scales, however, it becomes weak at small scales (Recent experiments show that the Newtonian gravitational force is led down to $0.13 mm \sim 0.16 mm$ \cite{ch1b+}). Meanwhile, it is hard to combine the Newtonian gravitational force with quantum mechanics. In Eq. (\ref{eq18}), it is clear that the GUP corrected Newton's law is dependent not only on the Newton's gravitational constant $G$, the mass of two bodies $M$ and $m$, and the distances $R$ but also on the dimensionless constants $\alpha_0$ and $\beta_0$ as well as the Planck length $\ell _p$. Therefore, due to the effect of GUP, the result shows that the Newtonian gravitational force is valid at scales which is smaller than the order of a millimeter. When  $\alpha_0=\beta_0 = 0$, the Eq. (\ref{eq18}) reduces to the original Newton's law. Moreover, one can obtain the Newtonian potential from Eq. (\ref{eq18})
\begin{equation}
\label{eq19}
V\left( R \right) =  - \frac{{GMm}}{R}\left\{ {1 + \left\{ {\frac{{\alpha _0^2 }}{{32R}}\left[ {\frac{1}{{32}} - \ln \left( {\frac{{\pi R}}{{\ell _p^2 }}} \right)} \right] - 2R\beta _0^2 \ln R} \right\}\ell _p^2  + {\cal O}\left( {\ell _p^4 } \right)} \right\}.
\end{equation}
It is interesting to compare Eq. (\ref{eq19}) with the predictions that came from higher order corrections to the Newtonian potential in the Randall-Sundrum II (RS II) \cite{ch37}, the modification in Newton's gravitational potential on brane is \cite{ch38}
\begin{equation}
\label{eq20}
V\left( R \right) \sim \left\{ {\begin{array}{*{20}c}
   {\begin{array}{*{20}c}
   {\frac{{GMm}}{r}\left( {1 + \frac{{4l_\mu  }}{{3\pi r}} - ...} \right)} & {for} & {l_\mu   \gg r}  \\
\end{array}}  \\
   {\begin{array}{*{20}c}
   {\frac{{GMm}}{r}\left( {1 + \frac{{2l_\mu  }}{{3\pi r^2 }} - ...} \right)} & {for} & {l_\mu   \ll r}  \\
\end{array}}  \\
\end{array}} \right.,
\end{equation}
where $r$  and $l_\mu$  are the radius and the characteristic length scale of the theory, respectively. Our result agrees with the Newtonian potential in RS II when  $l_\mu  \gg r$. Hence, it suggests that $\left( {\alpha _0 ,\beta _0 } \right) \sim l_\mu$ can help us to set a new upper bound on the dimensionless constants $\left( {\alpha _0 ,\beta _0 } \right)$. Besides, both Eq. (\ref{eq20}) and the correction terms in Eq. (\ref{eq19}) become susceptible at a short distance, they indicate that GUP and brane world may predict the similar phenomena.

\section{The Quantum  Corrections to The Einstein's Field Equation}
A lot of works predict that the Einstein's field equation can be derived from entropic force. In this section, we will further investigate the laws of gravity and extend it to the relativistic case, so that we can obtain the modified Einstein's field equation via the GUP corrected entropic force. According to the GUP corrected number of bits, the bit density on the screen can be expressed as
\begin{equation}
\label{eq21}
dN = \frac{1}{{ \ell _p^2 }}\left[ {1 - \left( {\frac{{\alpha _0^2 \pi }}{{4A}} + \frac{{A\beta _0^2 }}{\pi }} \right)\ell _p^2} + {\cal O}\left( {\ell _p^4 } \right) \right]dA,
\end{equation}
where $A$ represents the area of the holographic screen, , and we use the natural units $c=k_B =1$ in the equation above. Here we assume that the energy associated with the mass $M$  is divided over $N$. Moreover, due to the equipartition law, it is easy to find that each bit carries $T/2$  mass. Therefore, the total mass is
\begin{equation}
\label{eq22}
M = \frac{1}{2}\int_\mathcal{S} {TdN} ,
\end{equation}
where $\mathcal{S}$ is the holographic screen. The local temperature can be expressed as
\begin{equation}
\label{eq23}
T = \hbar {{e^\phi n^b \nabla _b \varphi } \mathord{\left/
 {\vphantom {{e^\varphi  n^b \nabla _b \varphi } {2\pi }}} \right.
 \kern-\nulldelimiterspace} {2\pi }},
\end{equation}
where  $e^\phi$ is the redshift factor as the local temperature is measured by an observer from infinity \cite{ch39}. Substituting Eq. (\ref{eq21}) and Eq. (\ref{eq23}) into Eq. (\ref{eq22}), one has
\begin{equation}
\label{eq24}
M = \frac{1}{{4\pi G}}\int_\mathcal{S} {e^\phi  \nabla \phi \left[ {1 - \left( {\frac{{\alpha _0^2 \pi }}{{4A}} + \frac{{A\beta _0^2 }}{\pi }} \right)\ell _p^2 } +{\cal O}\left( {\ell _p^4 } \right) \right]dA.}
\end{equation}
It is necessary to mention that the integral on the right side of the above equation represents the modified Komar mass (the original Komar mass contained inside a volume in a static curved space time is defined as  $M_K  = \frac{1}{{4\pi G}}\int_\mathcal{S} {e^\phi  \nabla \phi  dA}$) \cite{ch18,ch39,ch6a+}, hence Eq. (\ref{eq24}) is the modified Gauss law in general relativity. Using the Stokes theorem and the Killing equation  $\nabla ^a \nabla _a \xi ^b  =  - R_a^b \xi ^a$, the Komar mass in terms of the Killing vector $\xi ^a$  and Ricci tensor $R_{ab}$ can be rewritten as \cite{ch18,ch40}
\begin{equation}
\label{eq25}
M_K  = \frac{1}{{4\pi G}}\int_\Sigma  {R_{ab} n^a \xi ^b dV} .
\end{equation}
Therefore, the Eq. (\ref{eq24}) becomes
\begin{equation}
\label{eq26}
M = \frac{1}{{4\pi G}}\int_\Sigma  {R_{ab} n^a \xi ^b dV}  + \frac{1}{{4\pi G}}e^\phi  \nabla \phi \int_{\cal S} {\left[ { - \left( {\frac{{\alpha _0^2 \pi }}{{4A}} + \frac{{A\beta _0^2 }}{\pi }} \right)\ell _p^2 } +{\cal O}\left( {\ell _p^4 } \right) \right]} dA,
\end{equation}
where  $\Sigma$ is the three dimensional volume bounded by the holographic screen  $\mathcal{S}$, and its normal is $n^a$. In \cite{ch6a+}, the authors showed that the $M$ can be expressed in terms of the stress energy tensor $T_{a b}$
\begin{equation}
\label{eq27}
M = 2\int_\Sigma  {dV\left( {T_{\mu \nu }  - \frac{1}{2}Tg_{\mu \nu } } \right)} n^a \xi ^b.
\end{equation}
As a result, substituting Eq. (\ref{eq27}) into Eq. (\ref{eq26}), one yields
\begin{equation}
\label{eq28}
\int_\Sigma  {\left[ {R_{ab}  - 8\pi G\left( {T_{ab}  - \frac{1}{2}Tg_{ab} } \right)} \right]n^a \xi ^b dV}  = \ell _p^2 \int_\mathcal{S} {e^\phi  } \nabla \phi \left[ {\frac{{\alpha _0^2 \pi }}{{4A}} + \frac{{A\beta _0^2 }}{\pi }+ {\cal O}\left( {\ell _p^4 } \right)} \right]dA.
\end{equation}
With the help of  $F = -m e^\phi \nabla \phi$, one can obtains the GUP corrected Einstein's field equation by some manipulations
\begin{equation}
\label{eq29}
R_{ab}  = 8\pi G\left( {T_{ab}  - \frac{1}{2}Tg_{ab} } \right)\left[ {1 - \left( {\frac{{\alpha _0^2 \pi }}{{4A}} + \frac{{A\beta _0^2 }}{\pi }} \right)\frac{{\ell _p^2 }}{{2\pi }} + {\cal O}\left( {\ell _p^4 } \right) } \right],
\end{equation}
with the  area of the holographic screen $A$. Obviously, this field equation is  dependent not only on the geometry of the space time and energy-momentum tensor but also on the GUP terms. For large horizon area or   $\alpha _0  = \beta _0  = 0$, the modified Einstein's field equation reduces to the original case.

\section{The Quantum  Corrections to The Friedmann Equation}
In \cite{ch16,ch23,ch26,ch40,ch41,ch42,ch43}, people analyzed the Friedmann equation by using the entropic force. Hence, we will study the effect of the GUP arising from Eq. (\ref{eq1}) on the form of the Friedmann equation. In the homogeneous and isotropic spacetime, the Friedmann-Robertson-Walker (FRW) universe is described by the line element
\begin{equation}
\label{eq30}
ds^2  = h_{\mu \nu } dx^\mu  dx^\nu   + \tilde r^2 d\Omega ^2 ,
\end{equation}
where  $\tilde r = ra\left( t \right)$,  $x^\mu   = \left( {t,r} \right)$, $d\Omega ^2  = d\theta ^2  + \sin ^2 \theta d\varphi ^2$ is the metric of two dimensional unit sphere, and $h_{\mu \nu }  = {\rm{diag}}\left[ { - 1,{{a^2 } \mathord{\left/ {\vphantom {{a^2 } {\left( {1 - kr^2 } \right)}}} \right.
 \kern-\nulldelimiterspace} {\left( {1 - kr^2 } \right)}}} \right]$ is the two dimensional metric with  $\mu = \nu =0, 1$, $k$  is the spatial curvature constant, respectively. Using the relation  $h^{\mu \nu } \partial _\mu  \tilde r\partial _\nu  \tilde r = 0$, the dynamical apparent horizon of the FRW universe can be expressed as
\begin{equation}
\label{eq31}
\tilde r = ar = {1 \mathord{\left/
 {\vphantom {1 {\sqrt {H^2  + {k \mathord{\left/
 {\vphantom {k {a^2 }}} \right.
 \kern-\nulldelimiterspace} {a^2 }}} }}} \right.
 \kern-\nulldelimiterspace} {\sqrt {H^2  + {k \mathord{\left/
 {\vphantom {k {a^2 }}} \right.
 \kern-\nulldelimiterspace} {a^2 }}} }},
\end{equation}
where  $H = {{\dot a} \mathord{\left/ {\vphantom {{\dot a} a}} \right. \kern-\nulldelimiterspace} a}$ is the Hubble parameter. Now, supposing that the matter source in the FRW universe is a perfect fluid, which stress-energy tensor is
\begin{equation}
\label{eq32}
T_{\mu \nu }  = \left( {\rho  + p} \right)u_\mu  u_\nu   + pg_{\mu \nu },
\end{equation}
where $u_\mu$ is the four velocity of the fluid. The conservation law of energy-momentum leads to the following continuity equation
\begin{equation}
\label{eq33}
\dot \rho  + 3H\left( {\rho  + p} \right) = 0.
\end{equation}
In order to investigate the GUP corrected Friedmann equation, one should consider a compact spatial region  $V = \left( {{4 \mathord{\left/
 {\vphantom {4 3}} \right. \kern-\nulldelimiterspace} 3}} \right)\pi \tilde r^3$ with a compact boundary  $\Sigma  = 4\pi \tilde r^2$. By combining the Eq. (\ref{eq18}) with Newton's second law, one has
\begin{equation}
\label{eq34}
F = m\left( {\frac{{\partial ^2 \tilde r}}{{\partial t^2 }}} \right) = m\ddot ar =  - G\frac{{Mm}}{{\tilde{r}^2 }}\chi \left( {\ell _p ,\alpha _0 ,\beta _0 } \right),
\end{equation}
where $\chi \left( {\ell _p ,\alpha _0 ,\beta _0} \right) = 1 + \left[ {2\beta _0^2 \tilde r^2  + {{\alpha _0^2 {\rm{ln}}\left( {\frac{{\pi \tilde r^2 }}{{\ell _p^2 }}} \right)} \mathord{\left/ {\vphantom {{\alpha _0^2 {\rm{ln}}\left( {\frac{{\pi \tilde r^2 }}{{\ell _p^2 }}} \right)} {16\tilde r^2 }}} \right. \kern-\nulldelimiterspace} {16\tilde r^2 }}} \right]\ell _p^2 + {\cal O}\left( {\ell _p^4 } \right)$ and $m$ represents the test particle near the holographic screen, the higher order terms ${\cal O}\left( {\ell _p^4 } \right)$  can be ignored since the $\ell _p$ is a very small value. The total physical mass $M$  inside the volume  $\mathcal{V}$ can be defined as
\begin{equation}
\label{eq35}
M = \int_{\cal V} {dV\left( {T_{\mu \nu } u^\mu  u^\nu  } \right)}  = \frac{4}{3}\pi \tilde r^3 \rho,
\end{equation}
where $\rho  = {M \mathord{\left/ {\vphantom {M V}} \right. \kern-\nulldelimiterspace} V}$ is the energy density of the matter in the spatial region  $V$. Putting Eq. (\ref{eq34}) into Eq. (\ref{eq33}), one has the acceleration equation
\begin{equation}
\label{eq36}
\frac{{\ddot a}}{a} =  - \frac{4}{3}\pi G\rho \chi \left( {\ell _p ,\alpha _0 ,\beta _0 } \right).
\end{equation}
For deriving the Friedmann equation, it is necessary to use the active gravitational mass (or Tolman-Komar mass) $\mathcal{M}$  instead of the total mass $M$ because the acceleration in a dynamical background is produced by the active gravitational mass. According to \cite{ch23}, one can express the active gravitational mass in terms of energy-momentum tensor $T_{\mu \nu}$
\begin{equation}
\label{eq37}
\mathcal{M }= 2\int_\mathcal{V} {dV\left( {T_{\mu \nu }  - \frac{1}{2}Tg_{\mu \nu } } \right)} u^\mu  u^\nu   = \frac{4}{3}\pi \tilde{r}^3 \left( {\rho  + 3p} \right).
\end{equation}
Replacing the total mass  $M$ by the active gravitational mass  $\mathcal{M}$, the Eq. (\ref{eq35}) can be rewritten as
\begin{equation}
\label{eq38}
\frac{{\ddot a}}{a} =  - \frac{4}{3}\pi G\left( {\rho  + 3p} \right)\chi \left( {\ell _p ,\alpha _0 ,\beta _0 } \right).
\end{equation}
The above equation is the GUP corrected acceleration equation for the dynamical evolution of the FRW universe. Using the continuity equation Eq. (\ref{eq32}) and multiplying both sides of Eq. (37) with  $\dot{a}a$, then integrating it, the result is \cite{ch26}
\begin{equation}
\label{eq39}
\frac{d}{{dt}}\left( {\dot a^2 } \right) = \frac{{8\pi G}}{3}\left[ {\frac{d}{{dt}}\left( {\rho a^2 } \right)} \right]\chi \left( {\ell _p ,\alpha _0 ,\beta _0 } \right),
\end{equation}
Integrating both sides for each term of Eq. (\ref{eq38}), one has
\begin{equation}
\label{eq40}
\dot a^2  + k = \frac{{8\pi G}}{3}\rho a^2 \left\{ {1 + \frac{1}{{\rho a^2 }}\int {\left[ {2\left( {ra} \right)^2 \beta _0^2  + \frac{{\alpha _0^2 \pi }}{{16\pi \left( {ra} \right)^2 }}{\rm{ln}}\left( {\frac{{\pi \left( {ra} \right)^2 }}{{\ell _p^2 }}} \right)} \right]\ell _p^2 d\left( {\rho a^2 } \right)} } \right\},
\end{equation}
the above equation can be rewritten as
\begin{equation}
\label{eq41}
H^2  + \frac{k}{{a^2 }} = \frac{{8\pi G}}{3}\rho \left\{ {1 + \frac{1}{{\rho a^2 }}\int {\left[ {2\left( {ra} \right)^2 \beta _0^2  + \frac{{\alpha _0^2 \pi }}{{16\pi \left( {ra} \right)^2 }}{\rm{ln}}\left( {\frac{{\pi \left( {ra} \right)^2 }}{{\ell _p^2 }}} \right)} \right]\ell _p^2 d\left( {\rho a^2 } \right)} } \right\}.
\end{equation}
It should be noted that $k$ is the spatial curvature which takes the values  $-1,0,1$, and the values correspond to a close, flat and open FRW universe, respectively. For calculating the correction term of Eq. (\ref{eq40}), we assume an equation of state parameter is  $\omega  = {p \mathord{\left/ {\vphantom {p \rho }} \right. \kern-\nulldelimiterspace} \rho }$, where  $\omega$ is a constant independent of time (or redshift), so, integrating the continuity equation Eq. (\ref{eq33}), it yields
\begin{equation}
\label{eq42}
\rho  = \rho _0 a^{ - 3\left( {1 + \omega } \right)} ,
\end{equation}
where  $\rho _0$ is an integration constant. Substituting Eq. (\ref{eq41}) into Eq. (\ref{eq40}) and integrating, the result is given by
\begin{equation}
\label{eq43}
H^2  + \frac{k}{{a^2 }} = \frac{{8\pi G}}{3}\rho \left\{ {1 + \left( {1 + 3\omega } \right)\left[ {\frac{{\alpha _0^2 }}{{72\tilde r^2 \left( {1 + \omega } \right)^2 }} + \frac{{2\beta _0^2 \tilde r^2 }}{{3\omega  - 1}} + \frac{{\alpha _0^2 }}{{24\tilde r^2 \left( {1 + \omega } \right)}}{\rm{ln}}\left( {\frac{{\tilde r\sqrt \pi  }}{{\ell _p }}} \right)} \right]\ell _p^2 } \right\}.
\end{equation}
With the help of Eq. (\ref{eq31}), the above equation can be further rewritten as
\begin{eqnarray}
 \left( {H^2  + \frac{k}{{a^2 }}} \right)\left\{ {1 - \left( {1 + 3\omega } \right)\ell _p^2 } \right.\left[ {\frac{{\alpha _0^2 }}{{72\left( {1 + \omega } \right)^2 }}\left( {H^2  + \frac{k}{{a^2 }}} \right) + \frac{{2\beta _0^2 }}{{3\omega  - 1}}\left( {H^2  + \frac{k}{{a^2 }}} \right)^{ - 1} } \right.
\nonumber \\
 \left. {\left. { + \frac{{\alpha _0^2 }}{{24 \left( {1 + \omega } \right)}}\left( {H^2  + \frac{k}{{a^2 }}} \right){\rm{ln}}\left[ {\left( {H^2  + \frac{k}{{a^2 }}} \right)\frac{{\ell _p^2 }}{\pi }} \right]} \right]} \right\} &= &\frac{{8\pi G}}{3}\rho .
 \label{eq44}
\end{eqnarray}
The above equation is the GUP corrected Friedmann equation of the FRW universe, which derived from the entropic force. It should be noted that Eq. (\ref{eq43}) is not only determined by the Hubble paramrter  $H$, the expansion scale factor of universe  $a$, the spatial curvature  $k$ and the constant $\omega$  but also affected by the dimensionless constants  $\alpha_0$, $\beta_0$ and the Planck length  $\ell _p$. For the present universe, the Eq. (\ref{eq43}) is nothing but a usual Friedmann equation since the apparent horizon radius is very large. However, the correction terms make sense when the apparent horizon radius at a short scale. Hence one can use the GUP corrected Friedmann to investigate the early stage of the universe. Moreover, in \cite{ch26,ch26+}, the authors concluded that the impact of quantum corrections at the early stage of the universe can affect the inflation, so that people may detect those consequences by astronomical observation.

\section{Conclusions}
In this paper, we studied the quantum corrections to the entropic force via a new kind of GUP that admits a minimal length, a minimal momentum and a maximal momentum. Firstly, we derived the modified entropy-area law of a black hole. Then, using the the modification of entropy, the GUP corrected number of bits  $N$ was obtained. Subsequently, based on the GUP corrected number of bits and Verlinde's conjecture about the entropic force, the GUP corrected Newton's law of gravitation, Einstein's field equation as well as the Friedmann equation have been investigated. The results showed that the GUP corrected Newton's law of gravitation, the GUP corrected Einstein's field equation and the GUP corrected Friedmann equation are dependent on the quantum correction term $\alpha_0$, $\beta_0$ and the Planck length $\ell_p$. These results agree with the original cases at a larger scale. However, when the length approaches to the order of Planck scale, the corrected results depart from the original cases since the GUP effect becomes susceptible at a short scale. Besides, it found that the GUP corrected Newton's law of gravitation is working at the sub-$\mu m$ range, and it can predict the similar phenomenon as the Randall-Sundrum II modle, this can help us to set a new upper bound on \((\alpha_0,\beta_0)\). Moreover, we found that the GUP corrected Friedmann equation can help people to study the properties of the early universe, and the impact of GUP may be detected by the astronomical observation.

\vspace*{3.0ex}
{\bf Acknowledgements}
\vspace*{1.0ex}
The authors would like to acknowledge useful discussions with Karissa Liu and Lorraine Boyd, and especially the anonymous referee for useful suggestions and enlightening comments. This work is supported by the Natural Science Foundation of China (Grant No. 11573022).

\end{document}